\documentclass[a4paper,11pt]{article}
\pdfoutput=1 

\usepackage{jheppub} 

\usepackage[T1]{fontenc} 

\usepackage{graphicx}

\title{A priority based noise tolerant jet framework and algorithm}

\author{Daniel~Duffty}

\author{and Zack~Sullivan}
\affiliation{Department of Physics, Illinois Institute of Technology, Chicago, Illinois 60616-3793, USA}

\emailAdd{dufftyd@wit.edu}
\emailAdd{Zack.Sullivan@IIT.edu}

\abstract{
  We introduce a new framework for jet definitions called p-jets that
  matches the computational speed of the currently used anti-$k_T$ jet
  algorithm, but avoids combining much of the energy from background
  pileup events with signal jets.  As a first illustration of the
  p-jet framework, we compare the effectiveness of a p-jet algorithm
  to the anti-$k_T$ algorithm in reconstructing low energy jets from
  resonant $Z$ boson production and 50 pileup events.
}


\arxivnumber{1606.04497}

\begin{document}
\maketitle
\flushbottom


\section{Introduction}
\label{sec:intro}

\indent\indent
In the culmination of five decades of experimental searches, the Large
Hadron Collider (LHC) at CERN was successful in its runs at 7 TeV and
8 TeV in discovering a Higgs boson
\cite{Aad:2012tfa,Chatrchyan:2012ufa}, the last particle of the
standard model, and in making significant improvements to many
standard model measurements.  In the coming 13 and 14 TeV runs, the
LHC will further refine these measurements, and search for evidence of
physics beyond the standard model (BSM).  Existing limits from
searches for BSM particles suggest their production is rare, and if
they are found they will likely have one of two detection signatures:
either the particles will decay to a small number of well-measured
highly energetic objects, or decay to a large number of low energy QCD
jets through cascade decays.  The signature of low energy (40--60 GeV)
jets is a common signal for supersymmetric particles
\cite{Aad:2015lea} and standard model backgrounds alike.
Unfortunately, while highly energetic objects will be well understood,
low energy jets will be highly subject to interpretation based on
choice of jet algorithm, and on a background that will dominate low
energy objects at the LHC.

During the LHC's high energy runs, the number of proton-proton
interactions per beam crossing (pileup) will increase from around 20
in the 8 TeV run to 50--140 in the 13 and 14 TeV runs.  The large
number of simultaneous events will cause significant degradation of
both energy resolution and event distinction.  For high energy objects
the energy contribution from this pileup will be insignificant, but
the energy contribution from pileup to low energy objects will be on
the order of the energy contribution for the physics processes
themselves, resulting in a very low signal to noise ratio.  An example
of how severe this effect can be is illustrated in Ref.\
\cite{Calkins:2013ega}; that study concentrated on the effect of
pileup for top quark reconstruction, but the reconstruction of a $W$
boson from dijets is most alarming.  With 50 pileup events the
reconstructed width of a $W$ approximately doubles, and by 140 pileup
events (predicted at the end of a planned luminosity upgrade), the
$W\to$ dijet ``peak'' is nothing more than a falling spectrum.  While
relatively new to particle physics, this challenge of a large QCD
radiation background has long been a limiting factor for jet analyses
in heavy-ion collisions.  Existing jet algorithms are insufficient to
overcome the worst of these backgrounds, hence the need for a
qualitatively new framework for jet reconstruction.

In recognition of the coming challenges for jet reconstruction, two
paths have been actively pursued.  The first is a shift in
concentration to the production of high energy jets that would come
from the tails of production cross sections.  The search for jet
substructure \cite{Butterworth:2008iy,Ellis:2009su,Krohn:2009th} and
boosted jet algorithms \cite{Abdesselam:2010pt} show great promise
where applicable, but ignore the bulk of the production cross
sections, such as the thresholds of all known standard model
particles.  An alternate series of methods have been proposed to deal
with pileup.  These range from selective subtraction of energy in the
detector to only clustering energy in the tracker (charged tracks)
\cite{Cacciari:2014gra,Krohn:2013lba,Perloff:2012,ATLAS}.  These
proposals have some merit in reconstructing large numbers of events,
but they share one restriction: they require a statistically large
sample for their event-by-event misreconstructions to cancel.  This
reduces their effectiveness in reconstructing rare events with low
energy jets.  Most of the BSM processes physicists hope to observe
will not have enough events for this cancellation to occur.  The
result of this is a significant reduction of discovery potential and
sensitivity to precision standard model physics.

The difficulties with current jet definitions can be traced to the
origin of the concept of jet clustering.  Since the observation of
jets at PETRA \cite{Brandelik:1979bd}, a typical final state with QCD
radiation in high energy physics has consisted of a small number of
collimated, well isolated groups of hadrons. These groups of hadrons
are classified as jets, and various ``jet algorithms'' were developed
to accurately include as much of this energy in a given jet as
possible. These jet algorithms have evolved much over the years, but
the underlying assumption of isolated jets is still present; the
problem is that with 50--140 pileup events this assumption is no
longer valid.  This paper proposes a new jet framework designed to
work in an intrinsically noisy QCD environment.  Instead of
fully-reconstructed isolated clusters of QCD radiation being the
dominant feature of measurements, jets will be small peaks of highly
correlated energy sitting on a loosely correlated sea of background
noise.

The method of clustering we propose begins with the reminder that,
while QCD radiation does tend to peak in particular directions, in
hadron colliders it is part of a continuum that connects the beam
remnants to energetic regions in the detector.  Since detector
resolutions and pileup imply we cannot reconstruct all of the QCD
radiation spectrum from any given event, we include only a well
understood core of the energy that would compose a traditional jet,
and compensate for the expected energy outside the small core
included.  This philosophy embodies itself in the structure of our jet
algorithm, the p-jet algorithm; unlike the jet algorithms commonly
used in the LHC analyses, known as sequential recombination
algorithms, the p-jet algorithm invokes a threshold for inclusion of
energy into a cluster that increases as the energy of the jet
increases.  The ultimate goals of this algorithm are to be unique and
stable in the expected high background environment of a 14 TeV LHC,
and to maintain accurate reconstruction of states from a low number of
events.  In section \ref{sec:algorithm} we introduce the p-jet
framework, and in section \ref{sec:Z} demonstrate a proof of concept
with an explicit jet algorithm that can reconstruct resonant $Z$ boson
production in a 50 pileup event environment.

\section{The p-jet framework and algorithm}
\label{sec:algorithm}

\indent\indent
In most modern collider experiments, calorimeter towers are clustered
into jets using a few sequential recombination algorithms, where
objects are combined in order of the smallest measurement of a
``distance'' metric.  The distance metric is measured in the space of
pseudorapidity $\eta$ or rapidity, related to the angle relative to
the beam, and the azimuthal angle $\phi$.  Sequential recombination
algorithms are distinguished by their the energy weighting, using some
power of the transverse momentum $p_T$ or transverse energy $E_T$ to
prioritize which towers are combined first.

The typical distance measure for modern recombination algorithms is 
\begin{equation}
\label{eq:ktdist}
d_{ij} = \min(p^{2k}_{Ti},p^{2k}_{Tj}) \frac{\Delta R_{ij}}{R_{\max}} < p^{2k}_{Ti} \,,
\end{equation}
with
\begin{equation}
\label{eq:delR}
\Delta R_{ij} = \sqrt{\Delta \eta_{ij}^2 + \Delta \phi_{ij}^2} \,.
\end{equation}
The power $k$ embodies the choice to prioritize either high or low
energetic towers; $k=1$ clusters low energy objects first (the $k_T$
algorithm \cite{Ellis:1993tq}), and $k=-1$ clusters high energy
objects first (the anti-$k_T$ algorithm \cite{Cacciari:2008gp}). The
anti-$k_T$ algorithm has the ability to form nearly conical jets of
radius $R_{\max}$, and it is often used to reconstruct non-boosted
jets. The choice of $k=0$ is called the Cambridge-Aachen algorithm,
and is frequently used in boosted object algorithms ($p_T > 300$--500
GeV) and jet substructure studies
\cite{Butterworth:2008iy,Ellis:2009su,Krohn:2009th,Abdesselam:2010pt}.
In this paper we compare to the anti-$k_T$ algorithm for low energy
jets, as this is main choice currently in use by the ATLAS and CMS
physics groups.

Sequential recombination algorithms operate via the following procedure:
\begin{enumerate}
\item Identify all objects to be grouped.
\item Create a list of distances ($d_{ij}$) between all reconstructed
  objects $i$ and $j$.
\item Identify the smallest distance.
\item Combine objects with smallest distance into a cluster.
\item Recalculate all distances between objects, including the cluster
  as an object.
\item If a $\Delta R_{ij} < R_{\max}$, return to step 3, else
  clustering is finished.
\item Apply a jet energy correction to the clusters we now call jets.
\end{enumerate}

The anti-$k_T$ version of this algorithm presents a problem for
reconstruction in a high pileup environment. The jets it produces are
typically cones of radius $R_{\mathrm{max}}$.  This allows for simple
subtraction of pileup energy based on an estimate of the average noise
in the cone's area, but it is highly dependent on a well modeled
pileup environment in the region of space around the physical jet.

In any single event the subtraction behaves poorly for light jets,
since the pileup noise is a significant fraction of overall anti-$k_T$
jet energy.  In addition, the anti-$k_T$ algorithm's ``strength'' of
having well defined jet sizes does not match the physics of the QCD
radiation spectrum; the more energetic an anti-$k_T$ jet is, the more
likely it is to take up the entire $\pi R^2_{\max}$ phase space due to
the inverse energy weighting in $d_{ij}$.  A real high energy jet is
boosted such that most of its energy is concentrated in a small cone,
while low energy jets have a wide angular distribution of energy ---
the opposite of the shape of the anti-$k_T$ algorithm.

A new jet algorithm is needed to accurately deal with pileup.  This
algorithm should satisfy several key features:
\begin{enumerate}
\item It should include a significant fraction of the QCD radiation
  from the hard physics process of interest, while rejecting the bulk
  of the noise from pileup.
\item It should have a well understood theoretical basis for
  comparison between experimental objects and theory calculations.
  This includes retaining both infrared and collinear safety to allow
  quantities to be theoretically derived.
\item It should be tuned to the shape of the QCD radiation spectrum.
\item It should perform at least as well as anti-$k_T$ in jet
  reconstructions in low pileup, and outperform in high pileup environments.
\item It needs to work on an event-by-event basis, rather than require
  statistical cancellations to average out misreconstructions and
  pileup subtraction.
\end{enumerate}

The p-jet framework we present essentially replaces the distance
measure with a new ``priority'' measure $p_{ij}$, which can be understood as
the percentage of a jet's energy above the threshold of another jet
\begin{equation}
\label{eq:priority}
p_{ij} = \max \left( \frac{E_j}{E_i} -T( \Delta R_{ij} ) , 0 \right) ,
\end{equation}
where $E_i\ge E_j$, and the threshold function $T \in [0,1]$ specifies the
algorithm, and is chosen below.  This threshold function sets the
``p-jet'' algorithm apart from the anti-$k_T$ family, since it no
longer guarantees a nearby energy cluster will be absorbed; only
objects with energy above the threshold are considered for clustering.
This threshold is designed to assure that collinear jets are combined,
but non-collinear noise (from low-level radiation and pileup) is
ignored.  A proper choice of $T$ will also have the strength that it
roughly matches the QCD radiation spectrum.  The priority function
should be proportional to energy so that high energy jets will be
narrower than low energy jets, i.e., low energy objects away from the
main cluster will be ignored.  With the exception of the distance
function's replacement with a priority function, the recombination and
p-jet frameworks share a similar overall structure and speed.

The p-jet framework mirrors a recombination algorithm procedurally:
\begin{enumerate}
\item Identify all objects to be grouped.
\item Create priority lists $p_{ij}$ of all objects $i$ on all objects $j$.
\item Identify the largest priority (closest to 1).
\item Combine objects with largest priority into a cluster $a$.
\item Recalculate all \textit{changed} priorities, ($p_{ia}$,
  $p_{ai}$), where the cluster is treated as a new object.
\item If a non-zero priority exists, return to step 3. If all
  priorities are 0, proceed.
\item Apply a jet energy correction to final clusters we call jets.
\end{enumerate}

The above framework can be used to build an explicit jet algorithm
with a suitable choice of threshold function $T$.  The difference
between this framework and the sequential recombination algorithms is
the priority function, whose difference from an arbitrary distance
measure is its distance-dependent energy-based threshold for
inclusion.  As an explicit example we choose a threshold function
which is designed to approximate the dominant QCD dipole radiation
spectrum as closely as possible,
\begin{equation}
\label{eq:threshold}
T(\Delta R_{ij}) = 2 \sin \left( 4 \pi \frac{\Delta R_{ij}}{R_{\max}} \right) .
\end{equation}
Matching the exact QCD spectrum is not absolutely necessary, but an
approximation is desirable as it enhances the ratio of signal to
noise without losing too much of the original signal.  The general
guidelines for any threshold function are as follows: The threshold
must be 0 at the origin and be monotonically increasing or flat.  The
$k_T$ algorithm can be thought of as a limiting case of the p-jet
family, where we set the threshold function to be $T(\Delta R_{ij}) =
\Theta(\Delta R_{ij}-R_{\max})$, i.e., set the threshold to 0 within the
radius of $R_{\max}$.

Pileup will have an approximately isotropic distribution due to its
source (low energy proton-proton scatterings).  The loosely correlated
assortment of pileup in any given crossing can be subtracted from
anti-$k_T$ jets using a local area-based subtraction.  This is
typically calculated in anti-$k_T$ algorithms by introducing ``ghost''
particles across the detector with 0 energy and keeping track of which
jet the ghost is clustered into.  The area encapsulated by the ghosts
can be then used to estimate how much energy to subtract from the
anti-$k_T$ jets.  For a large number of events this \textit{average}
subtraction will be correct.  For a small number of events, such as
those in a discovery search, this average subtraction method will
reduce the jet energy resolution and lead to a broadening of any
potential mass peak.  

Formally, any nontrivial p-jet threshold function produces a zero-area
jet, because the pairwise addition of energies above the thresholds
combines isolated points with zero area support.  This implies an
area-based subtraction of pileup for p-jets is unnecessary.  Finite
detector resolutions, found for example in physical calorimeter
towers, reintroduce small effective areas to the objects that are
summed.  Hence, some area subtraction could be introduced, but we find
it is generally not helpful.  Because significantly less pileup is
included into the jets by construction, there is no need for a large
statistical average to cancel out background fluctuations.  This leads
to smaller jet energy corrections, faster reconstructions, and better
mass resolution in high pileup p-jet analyses than in corresponding
anti-$k_T$ analyses.

\section{Benchmark $Z$ boson reconstruction}
\label{sec:Z}

\indent\indent 
We benchmark the p-jet framework, and the threshold function given by
eq.\ \ref{eq:threshold}, by simulating reconstruction of resonant
production of $Z$ bosons decaying to dijets, and comparing to
anti-$k_T$ jets.  Although the purpose of p-jets is to perform well in
a high pileup environment, we consider reconstruction with a moderate 50
minimum-bias events on top of the dijet signal, as this is closer to
the current environment at the LHC.  The simulation uses MadEvent 4
\cite{Alwall:2007st} to produce $Z\rightarrow q\bar{q}$, and PYTHIA 6
\cite{Sjostrand:2006za} for showering, hadronization, and minimum-bias
event generation.  In order to compare similar size jets, all
reconstructions use $R_{\mathrm{max}}=0.5$.

We use tracking information to distinguish vertices resulting from the
Z decay from the pileup events.\footnote{In this version of the
  algorithm we make no use of tracking information to improve jet
  energy resolution or scale.  This is clearly an area for improvement
  in future implementations.}  To simulate the tracking information
available in a detector, the stable charged particles from
PYTHIA are identified with their original vertex.  These
tracked particles are then clustered normally, and the resultant jets
are then associated with their dominant vertices.  Although each jet
will contain particles from several vertices, only a subset of those
vertices are likely candidates.  We considered two metrics to
categorize the jets: the number of charged particles inside the jet,
or the energy from charged particles inside the jet.  Since particles
from pileup are not coming from hard interactions, and tend to be
lower in energy and widely distributed, we choose to assign jets to
the vertex that contributed the most charged energy to a given jet.
More energetic and tightly clustered groups of particles are likely
coming from the hard interaction, so they will add a significant amount
of energy to only a few jets.  This means that an energy-weighted
charged track count is better able to distinguish between jets from
the hard interaction and the minimum bias interactions, regardless of
the actual number of charged particles produced.

By construction we miss some of the hadrons which originate with the
hard process.  To correct for the missing energy (or remove residual
pileup energy) we apply a jet energy correction (JEC).  We
characterize the JEC for p-jets (and anti-$k_T$, for comparison
purposes) on a jet-by-jet basis using $Z+$jet balancing.  The JEC for
any algorithm can always be fit to a known resonant state, but we are
interested in the relative performance under the supposition that we
may not know what new state we are looking for.  To calculate the JEC
we simulate $pp \rightarrow Zj$ in MadEvent, hadronize in
PYTHIA, and then cluster.  The energy of the partons (from
MadEvent) and the corresponding jets (from the clustered algorithm)
are compared, and a correction curve is calculated to adjust the
energy of the jets back to that of the original parton.

One subtlety of the JEC is that it must be calculated for each pileup
environment.  We compare the performance of the algorithms below by
considering the energy contribution from pileup to the predominately
``real'' jets in three forms: Subtract the typical energy from pileup
from each cell in the detector before clustering, subtract the pileup
energy from each clustered jet based on its size, or allow the
energy-based JEC to do all energy adjustment.  If the JEC is left to
account for pileup by itself, the correction approaches -40\% for
anti-$k_T$, and the p-jet correction is slightly less. If a per-cell
subtraction is done before clustering, the JEC shrinks significantly.
With an area-based subtraction after clustering, the JEC shrinks to a
minimum for the options considered, but still remains negative for
most jets.

\begin{figure}
\centerline{
\includegraphics[width=0.8\columnwidth]{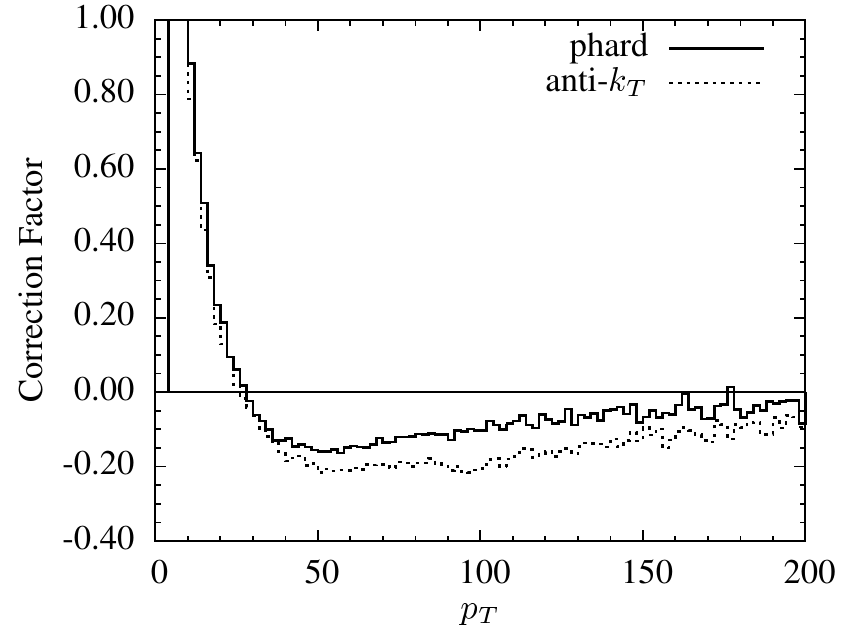}}
\caption{The fractional jet energy correction for p-jets and anti-$k_T$ jets
(using an area-based subtraction scheme) from $Z+j$ balancing with 50 events 
of pileup.
\label{fig:jec}}
\end{figure}

The jet energy correction required for p-jets is positive and greater
than for anti-$k_T$ jets before pileup is added.  This is due to
stricter requirements for energy inclusion in p-jets (proximity
\emph{and} energy rather than just proximity) than in anti-$k_T$ jets.
The JEC calculated in the presence of pileup is generally negative ---
very energetic jets cluster more energy from pileup than is lost by
the truncated size of the jet.  Due to its stricter requirements for
clustering, the p-jet correction is less severe (shown in figure
\ref{fig:jec}), in this case being much closer to 0, than the
correction for anti-$k_T$ jets, which is strongly negative, even after
an area based subtraction is implemented.  The relatively small JEC is
a sign that the p-jet framework is successful in its goals to
attenuate the effects of pileup, while retaining the primary jets.

\begin{figure}
\centerline{
\includegraphics[width=0.8\columnwidth]{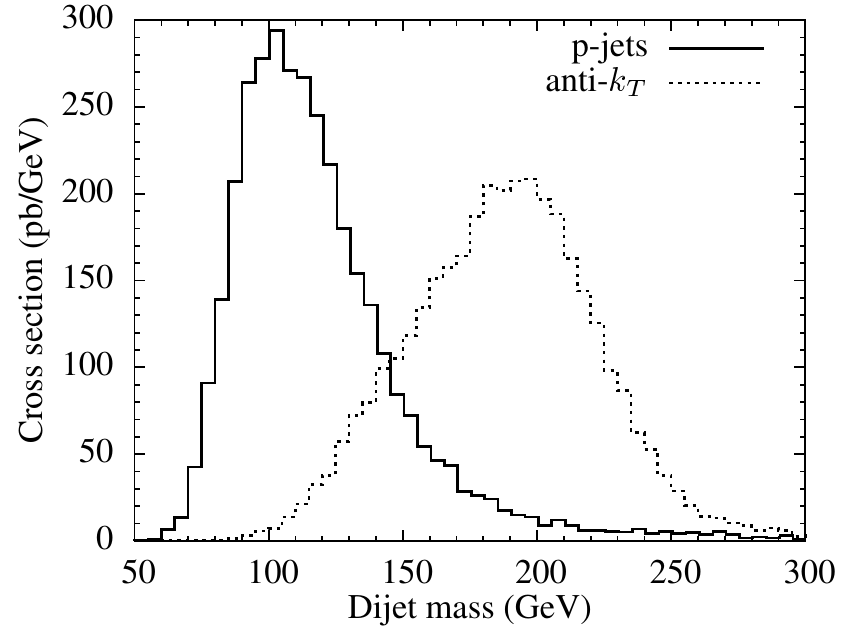}
}
\caption{The reconstructed dijet mass peak from $Z$ boson decay in 50 events
of pileup using p-jets and anti-$k_T$ jets after the jet energy correction.
\label{fig:ZJEC}}
\end{figure}

When we apply a jet energy correction, but do not do any per cell
energy subtraction, the p-jet algorithm performs significantly better
than anti-$k_T$ in reconstructing the position of the $Z\to$ dijet
invariant mass peak, and in energy resolution.  As we see in figure
\ref{fig:ZJEC}, the 102 GeV reconstructed mass from p-jets is slightly
higher than the $Z$ boson mass of 91 GeV, but slight shifts are
expected when using single-jet calibration to model dijet samples. The
jet energy resolution leads to a reconstructed width of 53 GeV.  The
anti-$k_T$ algorithm, by contrast reconstructs a central value of 188
GeV with a width of 82 GeV.  Current analyses typically perform
additional corrections to improve anti-$k_T$ jets, but the p-jet
algorithm is already performant.

\begin{figure}
\centerline{
\includegraphics[width=0.8\columnwidth]{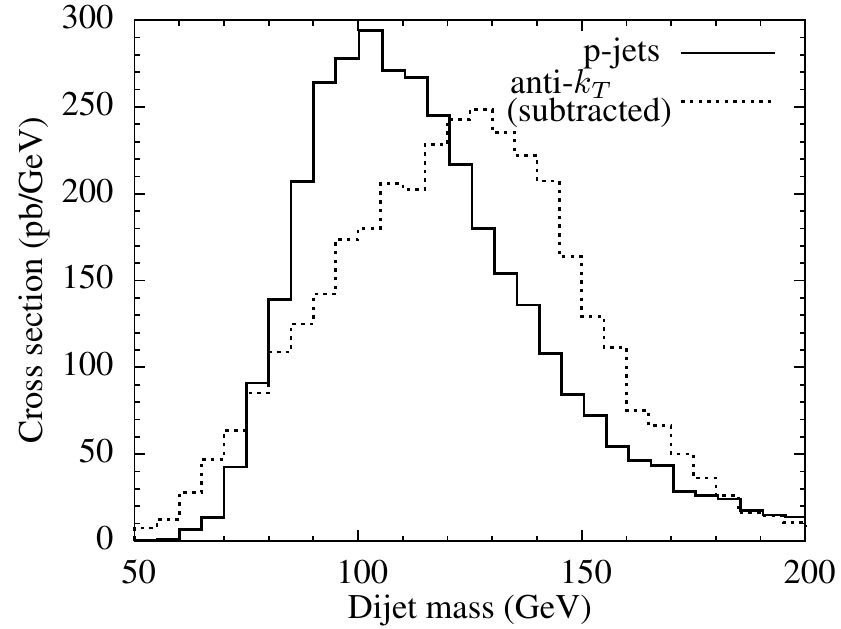}
}
\caption{The best reconstructed dijet mass peak from $Z$ boson decay in 50
events of pileup.  The p-jet reconstruction includes a JEC.  The anti-$k_T$
jets include both a JEC and a jet-area based energy subtraction.
\label{fig:Zpileup}}
\end{figure}

A fairer comparison to anti-$k_T$ involves additional subtractions
from the anti-$k_T$ jets tuned to the local jet environment
\cite{Cacciari:2014gra,Krohn:2013lba,Perloff:2012,ATLAS}.  If a
per-cell energy subtraction is performed before clustering, the
resultant anti-$k_T$ mass peak is reduced to 147 GeV with a width of
73 GeV, but has a long tail to higher energies.  P-jets perform
slightly worse with subtractions (105 GeV with a 60 GeV width), as is
expected since p-jets are formally zero-area jets --- there is no
energy$\times$area to subtract, so we are merely adding noise.  The
best $Z$ boson reconstruction for anti-$k_T$ jets is obtained by
performing an area-based subtraction after reconstruction.  In figure
\ref{fig:Zpileup} we see the best reconstruction of both algorithms,
where anti-$k_T$ reconstructed jets with area subtraction lead to a
mass measurement of 122 GeV and a width of 68 GeV.  This compares to
the simple JEC p-jet of 102 GeV and width of 53 GeV above.

While anti-$k_T$ jets can be further improved, our analysis
illustrates that without additional reclustering techniques, such as
pruning, p-jets outperforms anti-$k_T$ in the presence of pileup.  It
is tempting to think these techniques will produce jets equivalent to
p-jets given that, like p-jets, they remove soft energy from the jet.
However, recalling eq.\ \ref{eq:threshold}, the p-jet threshold
function we choose removes that energy predominantly from the tails of
the dipole radiation spectrum with only the softest radiation removed
near the core of the jet.  This represents the philosophy behind the
p-jet definition --- we cannot tell the difference between small
fluctuations and low-energy pileup, but those fluctuations are
expected by the real radiation spectrum.  Hence, we do not want to
artificially remove them as would be done by equivalently uniformly
raising the reconstruction energy threshold.

\begin{figure}
\centering
\includegraphics[width=0.45\textwidth]{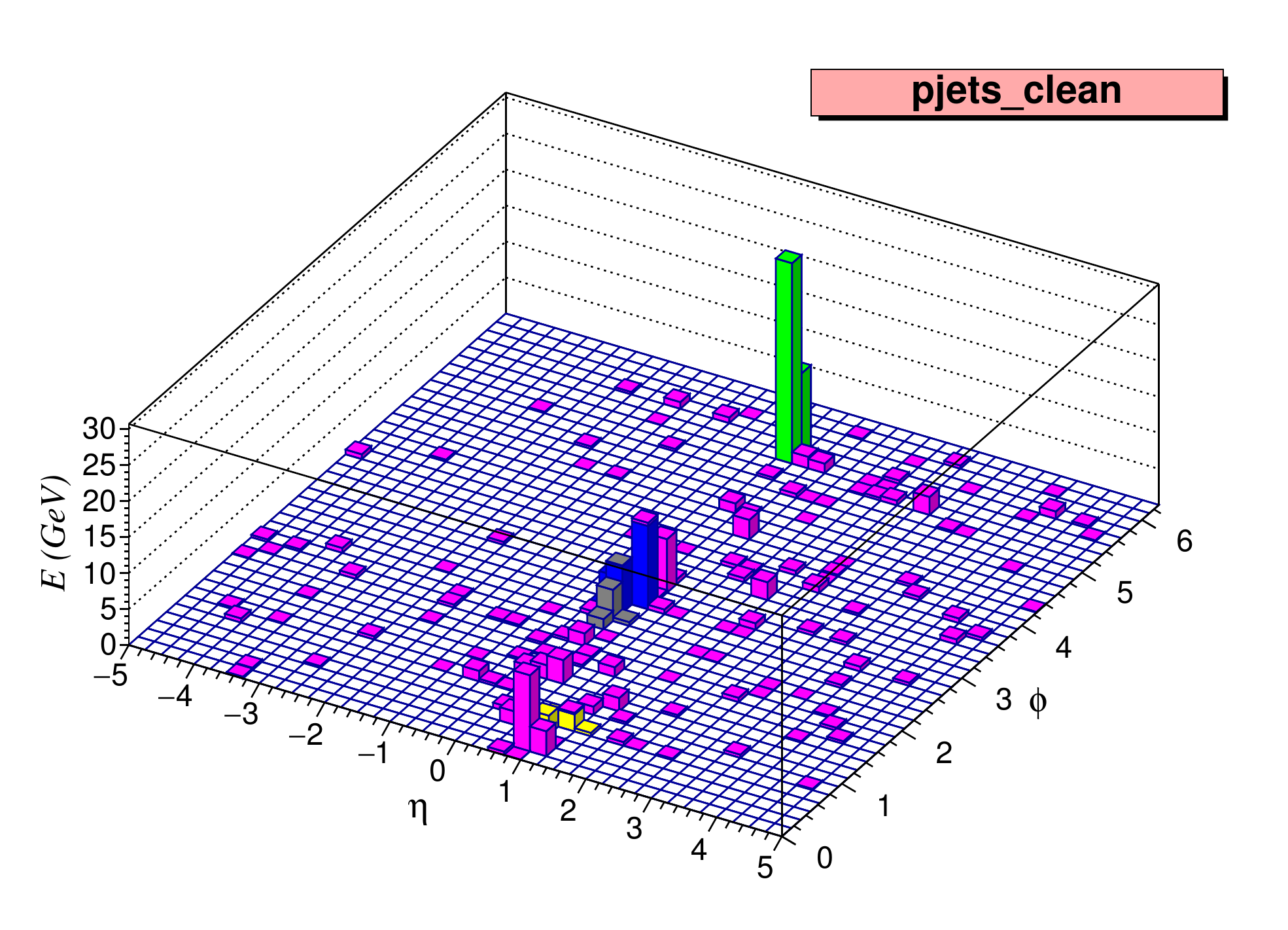}\ 
\includegraphics[width=0.45\textwidth]{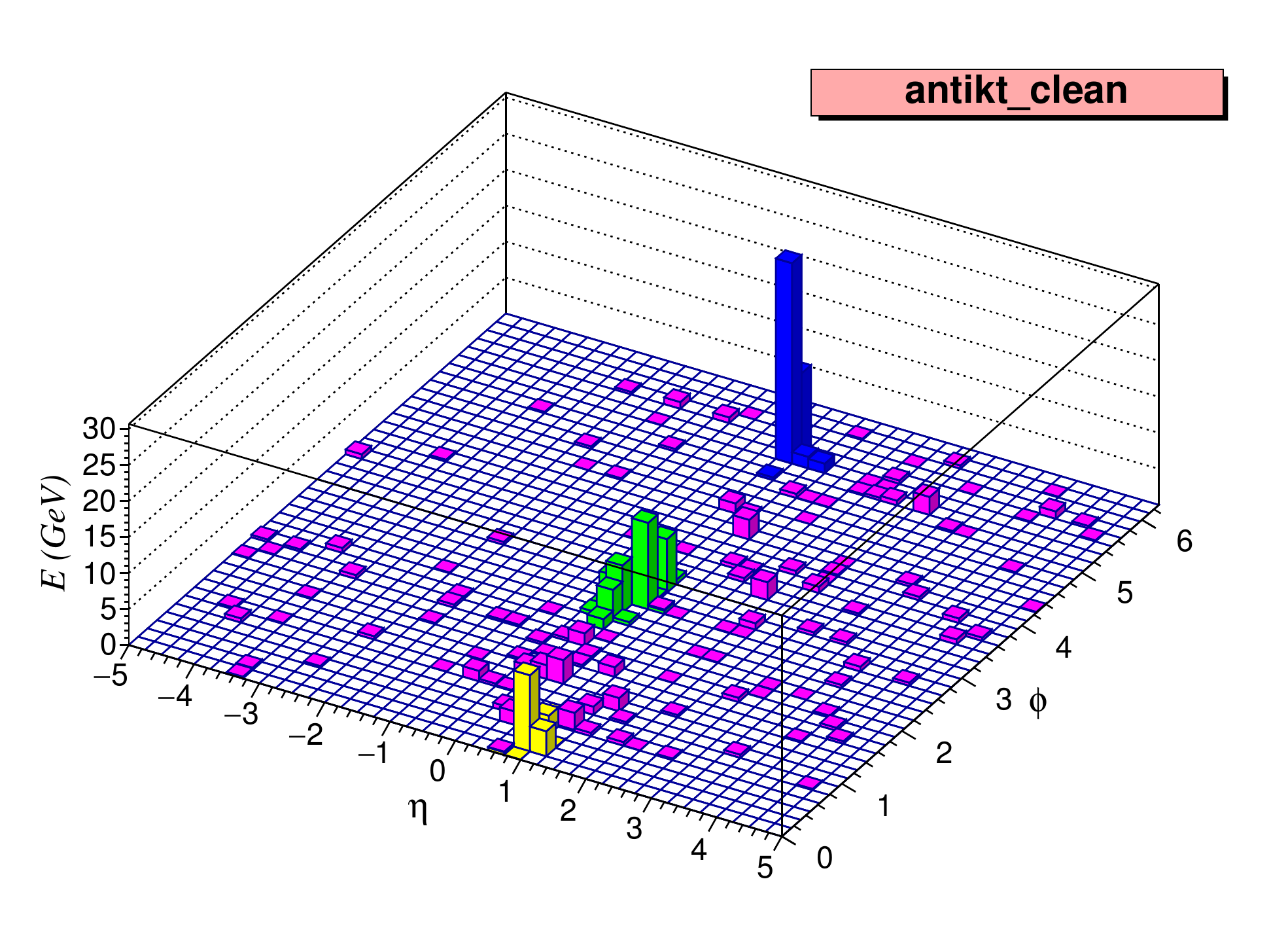}\\
\includegraphics[width=0.45\textwidth]{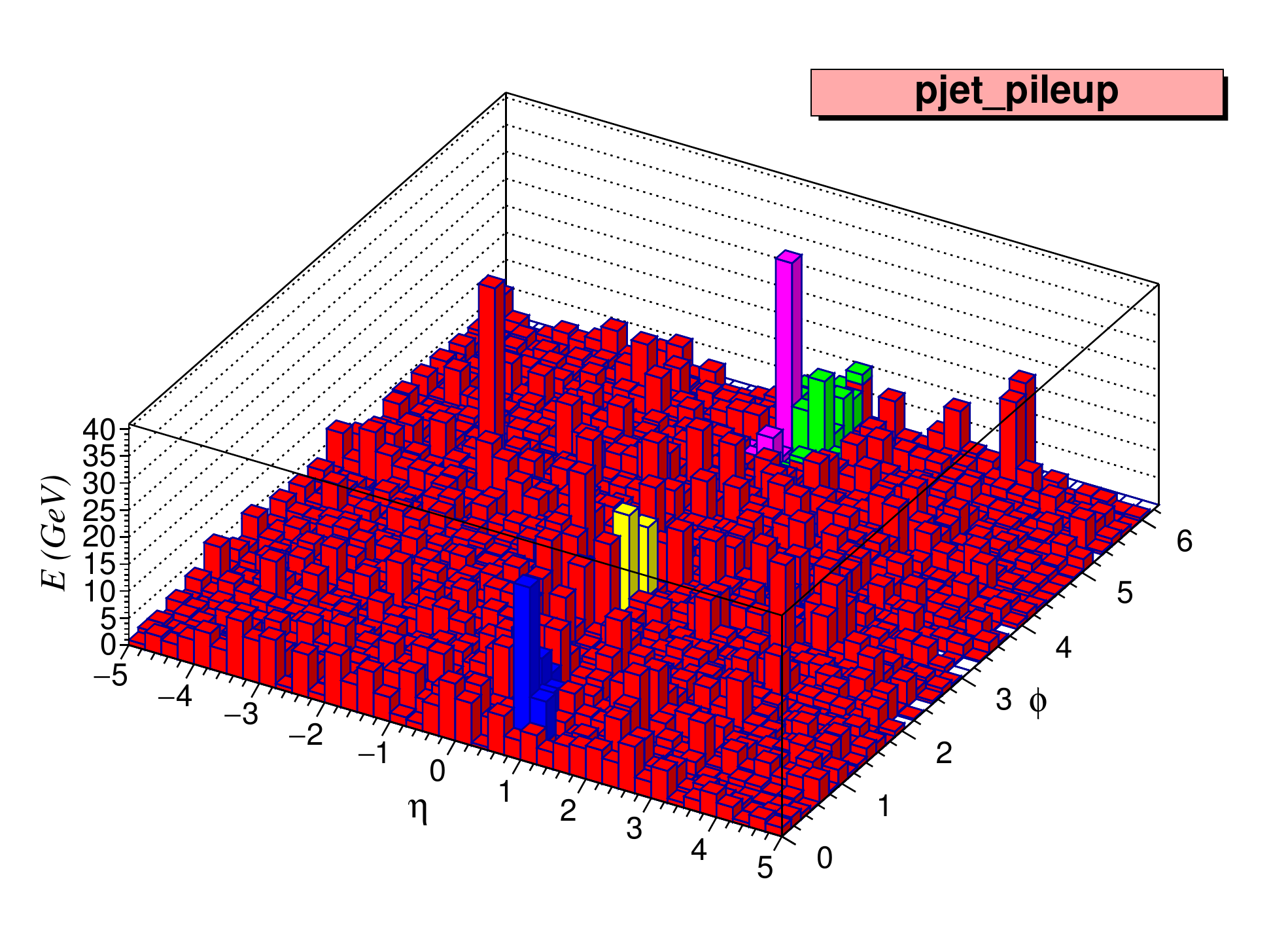}\ 
\includegraphics[width=0.45\textwidth]{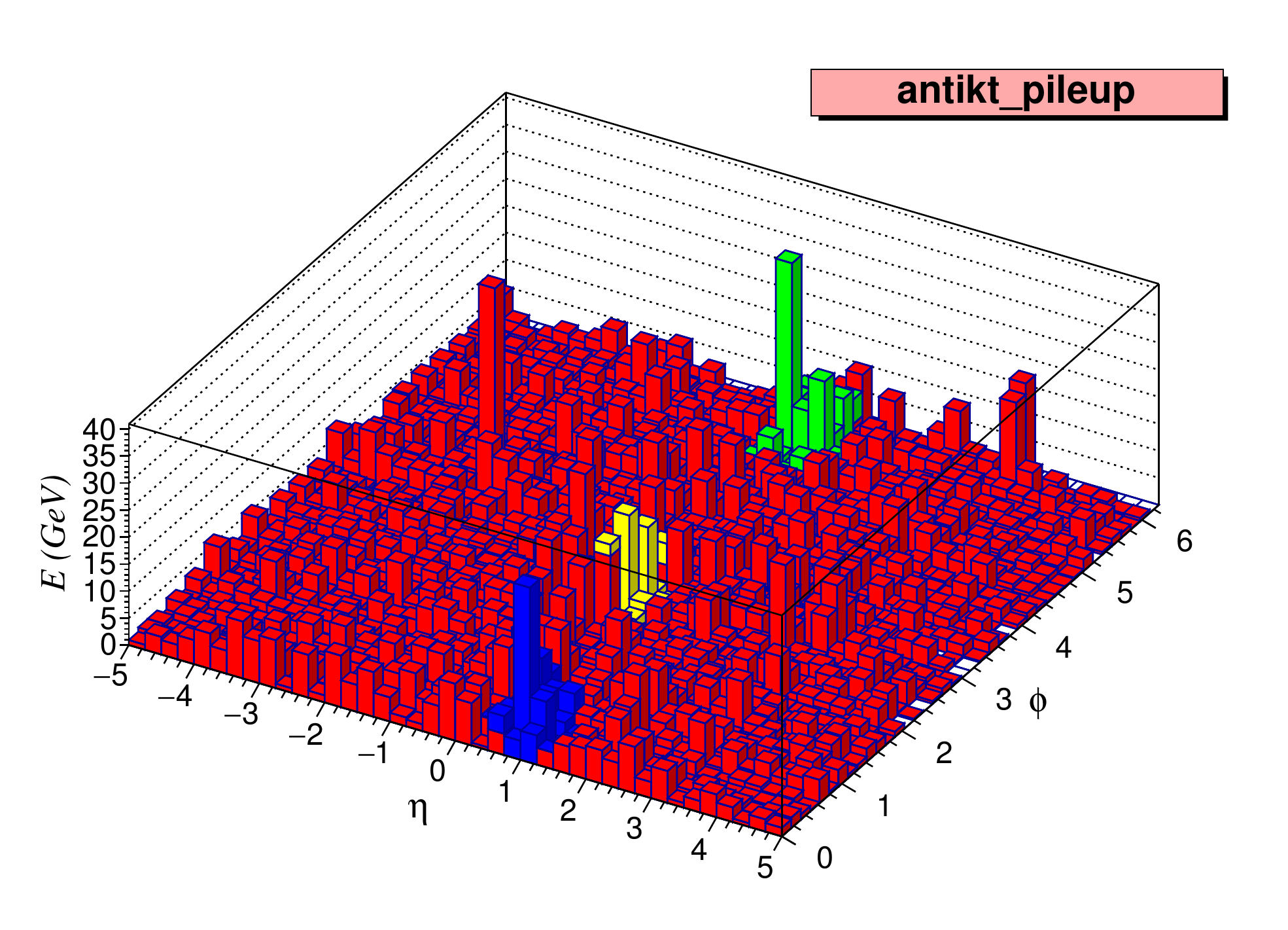}
\caption{A resonant $Z$ boson decay to dijets with initial state radiation
reconstructed with (left) p-jets and (right) anti-$k_T$ jets with (top) no
pileup or (bottom) 50 pileup events.
  \label{fig:lego}}
\end{figure}

For a simple channel, such as $Z\to$ dijets, both p-jets and
well-calibrated anti-$k_T$ can reconstruct to similar results.  The
algorithms, however, do lead to differences in detail at the level of
a single event.  We conclude this section by examining in figure
\ref{fig:lego} a single $Z$ decay with a hard initial state radiation
before and after pileup are included.  Before pileup we see that
p-jets and anti-$k_T$ jets reconstruct similar clusters, but
anti-$k_T$ swaps the order of the leading and second jet (when
compared to the ``truth'' level).  After pileup is added, both
algorithms find jets with the same relative energy ordering, but
clearly the p-jets are physically smaller.  As expected, anti-$k_T$
jets form compact cones of uniform size, regardless of the the
distribution of the energy in the event.  This is a reminder that
p-jets are much more like those that would be found by a $k_T$
algorithm.

The effectiveness of our choice of charged-particle energy vertex
association also appears in the pileup graphs of figure
\ref{fig:lego}.  E.g., the large tower at $\eta=-3$ is not associated
with our signal.  Other large energy towers are due to the random
overlap from multiple vertices, and are ignored by our clustering
technique.  Our the p-jet procedure is found to give a robust and
unique reconstruction, even in high density environments.

\section{Conclusions}
\label{sec:discussion}

\indent\indent 
We present a new priority-based framework, called ``p-jets,'' for the
reconstruction of jets in high luminosity environments.  Within the
p-jet framework, we examine one possible algorithm defined by its
threshold function $T$ designed to capture the dominant QCD dipole
shower.  As a proof of principle, we show that resonant $Z$ boson
production into dijets in high pileup is better reconstructed by
p-jets, than by the commonly used anti-$k_T$ algorithm unless
significant additional corrections are applied to anti-$k_T$ to
subtract noise absorbed within the jet cone.  Other than a jet energy
correction, which is applied to any jet, p-jets reach their potential
without additional subtractions as they are formally zero-area jets,
and avoid most pileup by construction.

The algorithm presented here is a starting point for p-jets.  Similar
quality of fit can be obtained for the modest luminosity examined here
using jet pruning and trimming techniques that recluster jets, rather
than avoiding noise absorption in the initial clustering.  P-jets
should be examined under a broad range of pileup scenarios, and for a
variety of jet threshold functions.  Ideally the choice of threshold
function would be optimized to improve signal extraction from physics
backgrounds.  For example, light quark or gluon-initiated jets tend to
have wider showers than heavy-quark initiated jets.  A threshold
function designed to reflect this could be used to improve $b$ or $c$
tagging.

The p-jet framework offers a new robust method for clustering QCD
radiation that is highly tolerant to noisy conditions, such as from
pileup in high luminosity colliders.  With a speed comparable to
existing algorithms, it should allow the capture of additional
structure on and event-by-event basis --- precisely what is needed for
rare BSM searches.


\acknowledgments

This work is supported by the U.S.\ Department of Energy under
Contract No.\ DE-SC0008347.  We thank Jesse Thaler for helpful
comments in clarifying the framework.

\appendix

\section{Small angle limit and infrared safety}
\label{sec:proof}

\indent\indent 
In QCD a jet algorithm removes singularities in a cross section
($\sigma$) calculation that arise in the perturbative calculation of
higher order radiation.  A simple example of this problem can be seen
in the three to two jet ratio arising in the process $e^+ e^-
\rightarrow \gamma \rightarrow q \bar{q} (+g)$.  
\begin{equation}
\label{eq:Xxsection}
\frac{\sigma_{q\bar{q}g}}{\sigma_{q\bar{q}}} = \int \frac{X_1^2+X_2^2}{(1-X_1)(1-X_2)},
\end{equation}
where $X_i={2 E_i}/{\sqrt{s}}$, and $\sqrt{s}$ is the center of
momentum energy of the process.  The $\frac{1}{1-X}$ terms produce
collinear singularities, since there is a direct relation between
$1-X$ and the angle $\theta_{jk}$ between two particles
\begin{equation}
\label{eq:XtoTheta}
(1-X_i) = X_j X_k (1-\cos(\theta_{jk})) \simeq \theta_{jk}^2 .
\end{equation}
An infrared singularity arises if the energy of the emitted gluon
approaches 0; e.g., if $X_3 \rightarrow 0$ (the gluon energy
approaches 0), $X_1 ,X_2 \rightarrow 1$, which again causes the
integral to be singular.
         
Both the infrared and collinear singularities are only an issue near
$X \rightarrow 1$.  The cross-section ratio can be made finite by
multiplying by a jet algorithm kernel that appropriately weights the
integral near $X=1$.  We choose to implement a weighting term of the
form $\theta^n$ (which translates to $(1-X)^n$) 
\begin{equation}
\label{eq:Jetxsection}
\int \frac{(X_1^2+X_2^2) \theta_{23}^{n \ge 0}}{1-X} = \int \frac{(X_1^2+X_2^2)}{(1-X_1)^{n < 1}} = \mathrm{finite}.
\end{equation}
For any $n>0$, the integral becomes finite as $X \rightarrow 1$ (which
is equivalent to $\theta \rightarrow 0$).  This weighting term is the
heart of the p-jet framework, as we have chosen to require only the
minimally sufficient condition to renormalize the singularities.  By
only specifying the endpoint limit, the p-jet framework provides
significant freedom in the choice of threshold function.  When combined
with the clustering procedure to produce a unique reconstruction this
freedom can be used to optimize the signal to background ratio in
noisy environments.

In order for a jet algorithm to be well defined, it must satisfy both
collinear and infrared safety.  We have illustrated how p-jets
satisfies this requirement from theory, but it must hold true for an
experiment as well.  Collinear safety arises as a stability issue when
the angle between two particles approaches zero, while infrared safety
refers to when the energy of the final state particles approaches
zero.  In practice, either of these singularities could cause
uniqueness issues from small fluctuations in jet energies resulting in
changes in the number of energetic reconstructed jets, which should be
constant despite small energy fluctuations.

A poorly designed jet algorithm would not immediately combine these
adjacent pairs, and possibly split what should be a single jet
combined into a single jet.  An example of where this type of failure
can occur is in a simple algorithm that absorbs all energy within the
$R_{\mathrm{cone}}$ radius of the most energetic tower.  If the energy
in the detector is instead split between two cells, the formerly
second most energetic single tower can become the most energetic and
seed the new jet causing a finite shift in the jet position.
Despite preferential clustering around the highest energy objects,
anti-$k_T$ avoids splitting problems because the $\Delta R_{ij}$ term
in the distance measure will approach 0 for nearly collinear jets
faster than the inverse energy weighting for any finite energy object.
In p-jets, $\Delta R_{ij} \rightarrow 0$ as well because the $T(\Delta R_{ij})$
will be 0 for collinear clusters and $p_{ij}$ will therefore be 1,
which is the maximum priority.

In experiment, infrared safety is an issue when it can cause jets to
merge or not due to small fluctuations in nearby energy.  Anti-$k_T$
jets avoid this issue by combining all minimal energy objects into the
highest energy cluster within range.  This causes the highest energy
jets to carve out a circular region in $\eta$--$\phi$ space from
nearby overlapping lower-energy jets.  The priority threshold function
$T(\Delta R_{ij})$ ensures a unique combination of objects as the
minimum value of all threshold functions evaluated at a point in
$\eta$--$\phi$ space identifies which cluster will be chosen to
combine with the object.


\end{document}